\documentclass[%
onecolumn,%
oneside,%
floats,%
aps,%
prd,%
nobibnotes,%
nofootinbib,%
amsmath,%
amssymb,%
amsfonts,%
amscd,%
superscriptaddress,%
]{revtex4}

\usepackage[utf8]{inputenc}
\usepackage{graphicx,array,dcolumn}
\usepackage[paperwidth=210mm,paperheight=297mm,centering,hmargin=1.8cm,vmargin=2.5cm]{geometry}

\def\larghezza{.36\textwidth}

\def\b{\textbf}
\def\i{\textit}
\def\mb{\mathbf}
\def\mc{\mathcal}
\def\be{\begin{eqnarray}}
\def\ee{\end{eqnarray}}
\def\mpl{m_{Pl}}
\def\sun{\odot}
\def\mn{{\mu\nu}}
\def\D{\mc D}
\def\-g{\sqrt{-g}}

\renewcommand\rho{\varrho}

\begin{document}
\def\larghezza{.36\textwidth}

\def\b{\textbf}
\def\i{\textit}
\def\mb{\mathbf}
\def\mc{\mathcal}
\def\be{\begin{eqnarray}}
\def\ee{\end{eqnarray}}
\def\mpl{m_{Pl}}
\def\sun{\odot}
\def\mn{{\mu\nu}}
\def\D{\mc D}
\def\-g{\sqrt{-g}}

\title{How to see an antistar}

\author{A.D. Dolgov}
\affiliation{Novosibirsk State University, Novosibirsk, 630090, Russia}
\affiliation{ A.I.~Alikhanov Institute of Theoretical and Experimental Physics, Moscow, 113259, Russia}
\affiliation{Dipartimento di Fisica e Scienze della Terra, Universit\`a degli Studi di Ferrara\\
Polo Scientifico e Tecnologico - Edificio C, Via Saragat 1, 44122 Ferrara, Italy}
\affiliation{Istituto Nazionale di Fisica Nucleare (INFN), Sezione di Ferrara\\
Polo Scientifico e Tecnologico - Edificio C, Via Saragat 1, 44122 Ferrara, Italy}

\author{V.~A.~Novikov }
\affiliation{ A.I.~Alikhanov Institute of Theoretical and Experimental Physics, Moscow, 113259, Russia}
\affiliation{Novosibirsk State University, Novosibirsk, 630090, Russia}
\affiliation{ Moscow Engineering Physics Institute, 115409, Moscow, Russia}
\affiliation{Moscow Institute of Physics and Technology, 141700, Dolgoprudny, Moscow Region, Russia}

\author{M.~I.~Vysotsky}
\affiliation{ A.I.~Alikhanov Institute of Theoretical and Experimental Physics, Moscow, 113259, Russia}
\affiliation{Novosibirsk State University, Novosibirsk, 630090, Russia}
\affiliation{ Moscow Engineering Physics Institute, 115409, Moscow, Russia}
\affiliation{Moscow Institute of Physics and Technology, 141700, Dolgoprudny, Moscow Region, Russia}
%\author{M.~I.~Vysotsky}
%\affiliation{ A.I.~Alikhanov Institute of Theoretical and Experimental Physics, Moscow, 113259, Russia}
%\affiliation{Novosibirsk State University, Novosibirsk, 630090, Russia}
%\affiliation{Moscow Engineering Physics Institute, 115409, Moscow, Russia}
%\affiliation{ Moscow Institute of Physics and Technology, 141700, Dolgoprudny, Moscow
%Region, Russia}

%\date{}

\begin{abstract}
 Polarization of photons emitted in weak decays occuring at distant star allows to determine
whether this star is made from antimatter. Even more promissing is the observation of
neutrinos (antineutrinos) produced at neutronization (antineutronization) reactions at the
beginning of SN ($\overline{SN}$) explosion.
\end{abstract}

\maketitle

According to the Standard Cosmological Model (SCM) no primordial antimatter
remains in the Universe. Let us shortly remind the arguments  which lead
to this conclusion. When in the course of the post Big Bang expansion
the universe cooled down below the QCD phase transition at
$T_{QCD} = 100 - 200 $ MeV, baryon-antibaryon pairs started  to annihilate.
If the baryon number of the Universe at these
temperatures was locally zero, then the remaining frozen concentration  of baryons
 would be  (see e.g. ref.~\cite{1}):
\begin{equation}
n_B/n_\gamma \approx 10^{-20} \, ,
\label{1}
\end{equation}
 where $n_B$ is the number density of baryons,  by assumption equal to that of
antibaryons, and $n_\gamma$ is the number density of photons in CMB.
This result
is by factor $10^{11}$ smaller than the presently observed baryon
concentration, which
can be e.g. deduced from the recent Planck
data~\cite{planck}, as:
\be
\eta =n_B / n_\gamma \approx 6 \times 10^{-10}\, ,
\label{beta}
\ee
with the precision at the per cent level.
Here it is implicitly assumed that the amount of antibaryons is negligibly small,
$n_{\bar B} \ll n_B$.

 In order to avoid conclusion (\ref{1}) we have either to assume that at the era of
baryon-antibaryon annihilation the universe was predominantly  and homogeneously populated
by baryons, or that the universe has domain structure with spatially separated
domains of matter and antimatter. In the last case it might be even not excluded that
the total baryonic number of the universe is zero.

 In the frameworks of the SCM the first option is accepted, which has a strong
support from the baryogenesis theory, whose basic principles have been formulated by
Sakharov almost half a century ago~\cite{sakharov}. In all known scenarios of baryogenesis
an excess of baryons over antibaryons was generated at very (or rather) high temperatures,
while at the subsequent cosmological expansion and cooling down the baryon-to-photon ratio
(\ref{beta})
stayed approximately constant, up to the entropy release by the massive particle
annihilation.

In the universe with an excess of baryons a chance for antibaryons to survive
was negligibly small, though in the early universe there were almost equal number
densities of baryons and antibaryons, $(n_B - n_{\bar B})/n_B  \approx \eta  \ll1$ .
The temperature at which the "massacre" of antibaryons by dominant baryons stopped
is fixed by the annihilation freezing which
 is determined by the time when  the annihilation rate became equal to the cosmological
 expansion rate:
\begin{equation}
\frac{1}{\sigma v \eta T_f^3}= \frac{M_p}{m_p T_f} \;\; ,
 \; T_f = m_p\sqrt{\frac{m_p}{100M_p \eta}} \approx 1 \; \mbox{\rm keV} \;\; ,
\label{3}
\end{equation}
where $\sigma v \approx 1/m_\pi^2$ is the cross-section of $p\bar p \to n\pi$ reaction
times the proton velocity in c.m. system, $m_p$ is the proton mass, $M_p$ is the Planck mass
and $n$ is the number of pions produced in $p\bar p$ annihilation.

That is why the remaining antiproton concentration being proportional to
\begin{equation}
\exp(-m_p/T_f) \sim 10^{-400000}
\label{4}
\end{equation}
is unobservably  small:  there is not a single primordial antiproton in all presently visible
part of our Universe. This bound is evidently too strong. Statistical fluctuations of the
antibaryonic density could strongly violate it but still the amount of the primordial
antiprotons would remain negligible.

The fluxes of the observed in cosmic rays antiprotons and positrons are about 4 orders
of magnitude smaller than the fluxes of protons and electrons respectively. They are
believed to be of secondary origin, produced in catastrophic processes in stars and in
interactions of energetic cosmic ray protons and electrons with matter.

 No  antinuclei  are observed in cosmic rays. The flux of secondary produced
anti-deuterium is  estimated as~\cite{anti-nucl}:
\be
F_{\bar D} \sim 10^{-7}{\rm  m^{-2} s^{-1} sr^{-1} (GeV/n)}^{-1} ,
\label{F-D}
\ee
where ${\rm Gev}/n$
is kinetic energy per nucleon. In other words, the predicted flux of $\bar D$
would be 5 orders of magnitude lower than the flux of antiprotons.
According to the estimates of the same paper~\cite{anti-nucl},
the fluxes of secondary produced $^3\bar He$ and $^4\bar He$ are respectively 4 and
8 orders of magnitude smaller than the flux of $\bar D$. A registration of antinuclei with the
flux above those predictions would be an unambiguous proof of existence of primordial
cosmic  antimatter.  An active search of cosmic antinuclei  was and is performed at several
balloon (BESS) and  satellite (AMS, PAMELA) missions. No single event of observation
of anti-helium or any heavier antinuclei  was reported.
The best up-to-date limit on the anti-helium flux was reported by BESS~\cite{bess} and it is:
$\bar He/He < 7\times10^{-8}$. Potentially PAMELA and AMS might improve this limit by an order
and two orders of magnitude respectively. There are some new projects with even larger
sensitivity.
 An observational limit on the flux of $\bar D$ was obtained in \cite{bess2} by far
above the theoretical expectation
of the secondary production flux.

An indirect signature of cosmic antimatter could be a flux of low energy, $\sim (0.1-1)$ GeV,
photons, originating from $\bar p p$--annihilation. There is no evidence of
any excess of such radiation which demands for its explanation the annihilation
source. So these data are used for quite restrictive limits on cosmologically
large clumps of antimatter. In particular, the nearest anti-galaxy could be at least at
the distance of 10 Mpc from us~\cite{anti-gal}. Similar considerations allow to conclude
that the fraction of antimatter in colliding galaxies of Bullet cluster is smaller than
$3\times 10^{-6}$~\cite{bullet}. As is was shown in ref.~\cite{cdg},
in baryo-symmetric universe with cosmologically large domains of matter and antimatter the
nearest domain of antimatter should be farther than a Gigaparsec away.

 All these bounds are applicable if matter and antimatter populate the universe in the similar
forms: clouds of gas and antigas, stars and antistars
of the same types, etc. However, it is possible to modify~\cite{ad-silk}
the baryogenesis scenario in such a way that antimatter would be mostly
hidden in compact stellar type objects, which could be in our Galaxy, even in
close vicinity to us. According to the suggested mechanism these objects were created
in very early universe, long before the recombination, and thus the usual CMB or LSS bounds
on antimatter are not directly applicable to them. These stellar-like (anti-)objects might
be abundant in the universe and even
make a noticeable contribution to the cosmological dark matter.

To make the paper self-contained we briefly present main features of such a model.
The starting point is the Affleck-Dine (AD) mechanism of baryogenesis~\cite{a-d},
where a scalar field $\chi$  with
non-zero baryonic number has the potential with flat directions. In the course of an early cosmological
evolution $\chi$ might acquire  a large expectation value along the flat direction and at a later
stage, when $\chi$ became massive its decay could create a large baryon asymmetry, $\eta$, which
in AD-model could be even of the order of unity. To make the scenario compatible with the data one
has to invent a mechanism to suppress $\eta$ down to the observed value. However, it is
possible to modify the AD-mechanism in a simple way, so that a large $\eta$ was generated only
in a small fraction of the total space. To realize such a picture it is sufficient to add  a general
renormalizable coupling of $\chi$ to the inflaton field $\Phi$:
\be
V(\chi,\Phi) = \lambda |\chi|^2 (\Phi - \Phi_1)^2 \; .
\label{V-chi-Phi}
\ee
In such a case the ``gates'' to the flat directions would be open only for
a short time when the inflaton field $\Phi$ was close to $\Phi_1$.
Hence the probability of the penetration to the flat
directions is small and $\chi$ could acquire a large expectation
value only in a tiny fraction of space. The universe would have
a homogeneous background of baryon asymmetry
$\eta \sim 6 \cdot 10^{-10}$ generated by the same field
$\chi$ which did not penetrate to larger distance through the
narrow gate or by another mechanism of baryogenesis, while the regions of
high density baryonic  matter, $\eta >0$, or antimatter, $\eta < 0$,
would be rare, but their contribution to
the total cosmological mass density might be significant or even
dominant. Let us call these
bubbles with high baryonic number density B-balls.

Originally the density contrast of B-balls with respect to the average cosmological
energy density
was very small (isocurvature fluctuations) but  after the QCD phase transition
such bubbles with large baryonic and/or  antibaryonic density would become much heavier
than the background medium (of the same volume), so they could form stellar-like astrophysical objects
at the very early stage of the cosmological history. As is shown in ref.~\cite{bambi-ad}
such antimatter  bubbles could survive in the early universe against annihilation with the background
of baryonic matter with small asymmetry $\eta \sim 10^{-9}$. Physically it is practically evident because
the mean free path of the particles of normal matter in the early universe is very short, so the
annihilation could  proceed only on the surface of the high-B bubbles, which has quite low efficiency.
At later cosmological stages the same reason prevents from strong annihilation again because of a short
mean free path inside such bubbles with high baryonic density.
These bubbles might form different types of astrophysical objects, from primordial black holes, compact
stars, e.g. similar to red giant cores or white dwarfs, or even resemble almost normal stars.
Their observational manifestations in the Galaxy,
such as e.g. an existence  of MeV photons from the annihilation $e^+ e^- \to 2\gamma$ and
more energetic $\gamma, e^-,$ and $e^+$ from $p \bar p$-annihilation,
were analyzed in ref.~\cite{bambi-ad}, where it was found that no data at the present time
are at odds with such a hypothesis.

In the present paper we wish to suggest  an alternative way to search for antistars in
galaxies through a difference of the polarization of radiation emitted by stars and
antistars (it will work for antigalaxies as well).

Usually it is supposed that in order to determine if the neighboring star is an antistar,
the phenomenon of CP-violation should be used. Just after the discovery of CP violation in
neutral kaons \cite{3} the following scenario was discussed: the inhabitants of the explored
star system were asked, if the shells of their atoms were made from the light charged leptons
which were more frequently produced in $K_L$ decays $K_L \to \pi^\pm e^\mp \nu$. If the answer
is ``yes'' -- then we are dealing with an antistar. The problem is that to realize such a scenario
we need to establish communication with the inhabitants of
another stellar system, which does not seem an easy task.

As it has been noted in \cite{4}, if such a scenario can be realized, it assumes communication
by radiowaves, so photons emitted on the Earth are detected and analyzed at the
stellar system under scrutiny.
But in this case CP-violation is not needed: we can send left-handed photons telling,
that polarization of charged lepton emitted in neutron $\beta$ decay is mainly the same. This
is the way to understand if the investigated system is made from antimatter.
But what  can we do if
the  stellar system is not inhabited or we are not able to establish a contact with inhabitants?

So the question we address is how one can distinguish  from
observations of a given star whether it is a normal star or an
antistar? If neutrinos produced in thermonuclear reactions are
detectable on the Earth, we will immediately find out, whether it
is a star which emits neutrinos, or an antistar which emits
antineutrinos. However the flux of neutrinos from stars is too low
to be detected: even the observation of neutrinos from the Sun is
highly nontrivial: the registered number of events is small. More
promising is a supernova explosion. It starts from neutronization
reaction $pe^- \to n\nu$, in which neutrinos are emitted. If
instead from the first stage of SN explosion antineutrinos are
detected on the Earth, it would mean that an antistar exploded
\cite{Weiler}. Let us mention that  detectors on the Earth waiting
for SN explosion in our galaxy are capable to detect neutrinos
from neutronization and distinguish them from antineutrinos. (Let
us note that spin-flavour conversion of Majorana neutrinos would
mimic $\overline{SN}$ explosion \cite{Ah}. However the spectra of
the detected antineutrinos should be different in this case and in
the case we consider).

The next question is what one can do if only photons emitted by a star are detected.
Usually one would think that the only way to distinguish a star from an antistar is provided by
CP violation. In particular CP violation leads to a difference in intensity of atomic lines
emitted by atoms and antiatoms. Though  the energies of emitted
photons and the total widths of atomic lines are  the same for atoms and antiatoms
due to  CPT-invarance, the violation of CP leads to different probabilities
of particular transitions in atoms and antiatoms. This way to determine if we are dealing
with antistar was suggested in \cite{5}. However since CP violation in atomic transitions is
very weak, it would be  interesting to find an alternative way to search for antistars.

This way is provided by ordinary weak interaction processes with photon emission.
These photons are longitudinally polarized.  They could be separated  from the
overall photon background if they have well defined energy being created e.g. in two
body decays.  If detected on the Earth  such photons would have
opposite polarization to that found for the laboratory produced photons,
it would mean that they were emitted by an antistar.

Presently beauty meson decays originating from the $b\to s\gamma$ penguin transition are widely
discussed (see \cite{6} and references therein).
Since a left-handed $s$-quark is produced in this
decay (the probability of a right-handed $s$-quark emission is suppressed as
$(m_s/m_b)^2 \sim 10^{-3}$) the emitted photon should be left-handed as well. Monoenergetic
photons emitted in $B\to K^\star\gamma$ transitions could be used
for search of antistars, if there are not two major
problems: first, one can hardly imagine production of $B$-mesons in stars since they are
too heavy; second, even if $b\bar b$ pair  is produced, then the beauty quarks
 would mainly
reside in $B$- and $\bar B$-mesons. Hence the photons
produced in $B\to K^\star \gamma$ and
$\bar B\to K^\star \gamma$ decays would have opposite polarizations.

The situation with strange quarks looks more optimistic.
First, lower energies are needed to produce strange particles.
Second, strange quarks would mainly reside in hyperons. {Photons} produced in
$\Sigma^+ \to p\gamma$ { decays have} large longitudinal polarization,
$\alpha = -0.76\pm 0.08$ \cite{7} and measuring their polarization at the Earth
we can  see if they were emitted by an antistar (in this case the
photon polarization is opposite).
Branching of  $\Sigma^+ \to p\gamma$ decay equals $(1.23 \pm 0.05)\times 10^{-3}$.
Stars with considerable amount of strange quarks are discussed in the literature
(the so-called strange stars \cite{8}).\footnote{We are grateful to A. Khodjamiryan
for this comment.} In outer shell of strange stars considerable amount of
$\Sigma$-hyperons should exist and studying the polarization of photons emitted in their
decays we can figure out if strange star is in fact an antistrange antistar.

The parity nonconservation in the $\gamma$-transitions of normal
nuclei made from protons and neutrons was observed in the Earth
experiments \cite{Lobashov}. The circular polarization of photons
appeared to be rather small: $P_\gamma = +(4 \pm 1) \cdot 10^{-5}$
for $~^{175}$Lu transition with the emission of 395 keV photon;
$P_\gamma = -(6\pm 1)\cdot 10^{-6}$ for 482 keV photon emitted in
$~^{181}$Ta transition and $P_\gamma = (1.9 \pm 0.3)\cdot 10^{-5}$
for the 1290 keV photon emitted in the $\gamma$-transition of
$~^{41}$K. The observation of the circular polarization of photons
with definite energy is ideally suited for the search of
antistars.

We are grateful to E. Akhmedov, S. Blinnikov, V. Egorychev, B.
Kerbikov and A. Khodjamiryan for useful discussions. The support
of the grant of the Russian Federation government 11.G34.31.0047.
is acknowledged. V.N. and M.V. are supported by the RFBR under the
Grants No. 11-02-00441, 12-02-00193 and by the Russian Federation
Government under Grant NSh-3172.2012.2 as well.

\end{document}